\documentstyle[12pt]{article}
\def\simlt{\lower.5ex\hbox{$\; \buildrel < \over \sim \;$}}
\def\simgt{\lower.5ex\hbox{$\; \buildrel > \over \sim \;$}}
\def\simpropto{\lower.2ex\hbox{$\; \buildrel \propto \over \sim \;$}}

\begin{document}
 \begin{center}
{\bf THE COSMIC MICROWAVE BACKGROUND}

\end{center}
\begin{center}{Joseph Silk\\
University of Oxford\\Astrophysics,
Denys Wilkinson Building,\\
Keble Road,
Oxford OX1 3RH, UK}

\end{center}

\begin{abstract}
I review the discovery of the temperature  fluctuations
 in the cosmic microwave background radiation. The underlying theory
and the implications for cosmology are reviewed, and  I describe the
prospects for future progress.
\end{abstract}

 \section{Introduction}
The discovery of  cosmic microwave background temperature fluctuations
in 1964 has revolutionised  cosmology.
Prior to 1992, one could only speculate about the initial conditions for
structure formation. Primordial acausal curvature fluctuations were first
detected by an experiment on the  COBE satellite 
that mapped the sky   on angular scales in excess of 5
degrees, much larger than the causal horizon of approximately one degree
at last scattering.
This provided the first glimpse of what was to come 
over the next ten years. Provided one adopted the
inflationary prescription of scale-invariant fluctuations, one  could
make the connection to the long sought seeds of large-scale structure
formation.
 But the connection was tenuous, and purely theoretical. 

There were two stages to the discovery. As early as 1967,
the prediction was made \cite{SacWol67}
 that intrinsic horizon-scale density  fluctuations could be as large as
10 percent without exceeding the current limits on the  cosmic
 microwave background isotropy at
that time. These fluctuations were considered to be  on scales
in excess of the last scattering horizon and were imprinted 
acausally at the
beginning of the universe. The connection with galaxy formation
arising
 from primordial
irregularities of infinitesimal amplitude, the current paradigm for
structure formation, was  also  made in
1967 \cite{s67}. The relic temperature fluctuations
 are  causally generated from  sound waves in
the primordial baryon-photon plasma on
the last scattering
surface
of the cosmic microwave background at redshift  $z\sim 1000.$

The search for
cosmic microwave background temperature fluctuations was long.
 Results from the COBE   far infrared spectrophotometer
(FIRAS) showed in 1990 that the spectrum of the   cosmic
 microwave background  did not deviate from
a blackbody by more than   $0.01\%$ of its  peak brightness.
  Large-angular scale fluctuations in the CMB were
eventually discovered in 1992 
using  the COBE differential microwave radiometer (DMR) \cite{cobe}
at a level $\delta T/T \sim 10^{-5},$ to within a factor of two of the
expectation from the twin  hypotheses of structure formation by
gravitational instability and a scale-invariant spectrum of primordial
adiabatic density fluctuations. 
However, the extrapolation from the COBE observations 
 to  scales of direct interest to
structure formation, from $\ell \sim 20$ to $\ell \sim 200$,
where the spherical harmonic wave-number $\ell\approx
180^\circ/\theta$
for angular scale $\theta,$
 was large.
It took nearly  a decade after the COBE launch before 
direct
detection of the elusive fluctuations that gave rise to large-scale
structure was  achieved. 

Sub-degree scale temperature
fluctuations from the surface of last scattering were finally
confirmed, and in particular the first  peak was measured,
by a series of balloon and mountain-top experiments 
\cite{miller}, \cite{debernadis}, \cite{hanany},
\cite{padin},\cite{halverson}, \cite{taylor},
\cite{archeops}, that spanned the range $\ell \sim 20$ to $\ell \sim 2000.$
 The
causal adiabatic fluctuations
were amplified from  primordial curvature fluctuations by subhorizon
growth in the matter-dominated era. This enhancement was detected as a
series of so-called  acoustic peaks, the first and best-measured
feature
  (known to better
than
ten standard deviations by the end of 2002) corresponding to the
angular scale
 of the
horizon at last scattering.  The peaks are driven by sound waves in
the primordial
baryon-photon plasma, and the sound wave amplitude  measures 
the seeds of present-day structure formation.

The theory of gravitational instability  in the expanding universe
as the driver of growth of structure from infinitesimal primordial
density
fluctuations,
pioneered in a remarkable  paper dating from  1946 \cite {lifshitz},
 was dramatically confirmed.
Gravitational instability  accounts for  the origin of large-scale
structure and in particular for the formation of the galaxies.    The
primordial structure was inhomogeneous  at a level of  around one part
in 100,000. On comoving scales in excess of 100 Mpc, the universe is
still uniform today and approaches  this level on the horizon scale. 
 However subhorizon growth means
that  the comoving scales of  galaxy clusters (approximately 10 Mpc) were first
nonlinear at a redshift of order unity, massive galaxies at  a
redshift of around 3, and dwarf galaxies at a redshift of order 10.
These inferences come from measurement of the power spectrum of
temperature fluctuations, from which the density fluctuation spectrum
can be recovered. The discovery of the fossil fluctuations that seeded
structure formation is a stunning intellectual achievement of the 20th
century.   

Complementary information comes from deep galaxy redshift
surveys and from the line-of-sight 
correlations of intergalactic clouds viewed via
Lyman alpha absorption against high redshift quasars.
It is  remarkable that the fluctuation spectrum measured at a
redshift of 1000 agrees in shape and normalization  with independent  
measurements of large-scale structure \cite{wss}, \cite{gawiser}.
These are performed 
at redshifts  of 0.3 and 3, corresponding to the typical
depths of the 2DF/SDSS
galaxy samples and the  
Lyman alpha forest  \cite {tegmark}. 
Even the bias
factor, which measures the ratio of the variance in galaxy counts to
dark matter density fluctuations, can now be inferred via 
a combination of any two of several 
independent
techniques: large-scale clustering \cite{peacock},  CMB data \cite{lahav},
  weak lensing \cite{brown}, \cite{hoekstra},
and the galaxy cluster abundance \cite{schuecker}. 
A bias of approximately 4/3 is inferred on galaxy cluster  scales.
This means that the rms fluctuations in dark matter density are
 75 percent  of the variance in luminous galaxy counts sampled over
randomly placed spheres of 10 Mpc diameter.
Galaxy formation is only weakly biased. 

\section{Some history} 

The cosmic microwave background and the temperature fluctuations
therein are two of the major predictions of the Big Bang theory. Others
are the expansion of the universe and the light element
abundances. The latter  motivated the cosmic microwave
background prediction.   Friedmann and independently Lema\^itre are
credited with the first predictions of the expansion of space. 
Indeed  Lema\^itre predicted a linear relation between galaxy redshift
and distance, verified by Hubble in 1929. Curiously, Hubble never
accepted the   Big Bang  interpretation of his redshift law as the
expansion of space.  It has even been speculated that Hubble's reluctance 
to accept that the eponymous law was a fundamental contribution to
the confirmation  of the theory of the expanding universe may explain
why  Nobel recognition eluded him.

Gamow pioneered the predictions of  element nucleosynthesis in the
first few minutes of the Big Bang \cite{gamow}, but
 the important recognition
by Hayashi in 1950
that the neutron abundance was determined by thermal
equilibrium in the primordial plasma set the scene for the later
precise predictions of the light element abundances.
Gamow's associates Alpher and
Herman \cite{alpher}
realized in 1949 that the inference the universe 
necessarily was hot at the epoch of  primordial nucleosynthesis
implied that, in order 
to get a reasonable helium abundance,  the
present epoch radiation background was characterised by a temperature
of 5 K. However they failed in any published work over the following
decade  to make the
connection with an observable microwave background. Indeed, later
papers speculated about a radiation temperature today as high as 50K.

The connection with observations  was first
attempted  in discussions in 1962 by Zeldovich \cite{zeld}
and in  1964 
by Doroshkevich and Novikov \cite{dorosh}, who even 
interpreted a Bell Labs measurement  
of the atmospheric temperature as setting
a limit of only 1 K  on the cosmic radiation background to be 
evidence for an initially cold universe. But in effect an
extraterrestrial  signal was
lurking in the measured antenna temperature.
Evidence of absence did
not  guarantee absence of evidence,  and the cosmic microwave
background was  discovered within a year via 
the epochal detection by Penzias and Wilson of what they described as
an excess antenna temperature, isotropic in the sky. Penzias and
Wilson did not initially realize the significance of their discovery,
and the connection with the Big Bang was made in a 
pioneering companion paper by
Dicke and collaborators at Princeton,
 who had been searching, simultaneously and independently, for 
the cosmic microwave background. The Princeton group realized that the
early universe would have been an ideal furnace for generating
blackbody radiation.
Dicke in particular, an experimentalist who was apparently
unaware of Gamow's work and was convinced that the Big Bang was
cyclic,
had long realized that light element synthesis in the Big Bang implied a
present day blackbody radiation field at a few K, peaking at microwave
frequencies. 

Observations of the blackbody-like 
spectral dependence of the CMB accumulated over
the next two and a half decades, but it was not until the launch of
the COBE FIRAS  interferometer that a definitive spectral
measurement was forthcoming. FIRAS confirmed the CMB spectrum was that
of a 
blackbody at $ 2.725 \pm 0.002 \rm K,$ with a precision
limited only by that of the on-board calibration source \cite{mather}.
The cosmic blackbody spectrum is
comparable to that of the best laboratory blackbody spectral
determination.
 
Parallel theoretical developments focussed on angular variations in
the CMB temperature. In 1967, Sachs and Wolfe had predicted that
primordial curvature fluctuations could exist on horizon  scales,
$\sim 1000$ Mpc,
generating temperature fluctuations
on large angular scales with an amplitude of order 1 percent. 
These would have been
acausal and attributable to the initial conditions of the Big Bang.
Subsequent studies of  the first instants of the universe
led to the development in 1981 of inflationary cosmology, which
predicted  adiabatic, gaussian fluctuations that were
boosted from quantum fluctuations to macroscopic scales during an
 early de Sitter phase.
The universe expanded exponentially during a brief moment, triggered
by the spontaneous breaking of the GUT symmetry at a temperature of
about $10^{16}$ GeV. The theory also predicted that the  fluctuations
should be approximately scale-invariant, although the amplitude was not
specified, in agreement with earlier conjectures
that have become known as the   Harrison--Zeldovich spectrum 
(\cite{harrison}, \cite{PY},  \cite{zeldovich}).
Many versions of inflation also subsequently predicted this, 
but perhaps the most noteworthy prediction of
inflation  is that the
 universe should be flat (\cite{guth},  \cite{albrecht},  \cite{linde}). 

Another pre-inflationary  prediction, and one that inflation 
did not, and could not, provide, 
was  that of the strength of the primordial
density
fluctuations. This yielded the first quantitative estimate 
of the fluctuations that are  required to have seeded structure  by
gravitational instability-induced growth. A simple prediction based on
adiabatic fluctuations required $\delta T/T$ to be a few parts in 
ten thousand at last scattering on angular scales of a few arc
minutes
in order to have formed galaxies and  clusters by the
 present epoch \cite{s67}.
The prediction simply utilized the adiabatic relation that, if
recombination were instantaneous, would give $\delta T/T\sim \frac{1}{3}
\delta\rho/\rho$.  

The current  prediction  is about  a factor of 10  smaller than implied by 
this simple expression.
What causes the diminution?
The realization that decoupling of the two fluids was not
instantaneous \cite{pee68} results in  a substantial
decrease of the relic temperature fluctuations \cite{sunyaev}.
Modern  treatments, which
incorporated  cold dark
matter, were first performed in
1984 (\cite{BonEfs84}, \cite{VitSil84}), and allow
fluctuation growth after the epoch of
matter-radiation equality but prior to last scattering, 
 anticipating a peak value $\delta T/T\sim 
3\times 10^{-5}$ on  an angular scale of about half a degree. This
corresponds to the horizon scale at last scattering.

The predictions refer to causal fluctuations generated by
gravitational instability on subhorizon scales during the era of
matter domination, and apply on the largest scales where structure is
seen, namely galaxy clusters and superclusters, and hence to angular
scales of order tens of arc minutes. The coherent peaks in the
radiation density are due to compressions and rarefactions of sound
waves in the primordial plasma as viewed at last scattering.  The
occurrence of multiple peaks in the radiation power spectrum was first
described in 1976 \cite{DSZ}, 
and were only accurately described as a generic description developed
from a multipole expansion of the CMB sky in 1981 \cite{SilWil80}.
This approach incorporated the first application to temperature
fluctuation predictions of the coupling of the primordial
photon-baryon plasma by the Boltzmann equation, originally given in a
classic paper in 1970 \cite{PY}. The latter paper indeed reported the
existence of coherent baryon oscillations of the adiabatic mode in the
matter-radiation plasma at matter-radiation decoupling, also
independently noted  by \cite{sunyaev}.  These two
papers quantify
 the so-called Sakharov oscillations \cite{sak} in the matter power
spectrum.  These matter power spectrum oscillations have
recently been detected, albeit with modest significance, in the 2DF
galaxy survey \cite{percival}.

Progressively more detailed
discussions in the 1990s describe a series of acoustic peaks,
interpreted
in the context of 
inflationary amplification of quantum fluctuations. These  
result in temporally coherent  fluctuations on the surface
of last scattering at $z\sim 1000$  that are driven by the growing
adiabatic mode of linear theory. 

The connection between $\delta T/T$ and structure formation presented a
definitive experimental target in the 1970s and 1980s, although
the goalposts were moved in 1984 when cold dark matter was first
included into the computational recipe. The prevalence of 
weakly interacting dark matter
that dominates the baryon density 
by a factor $\sim 10$  means that substantial  fluctuation growth
occurs between the epoch  of the onset of matter domination
at $z\sim 4\times 10^3,$ prior to last scattering at  $z\sim 10^3.$
This reduces the predicted fluctuation strength by a factor $\sim 4,$
and 
significantly modifies the detailed structure of the radiation power spectrum.

Upper limits were successively
improved. Only with the COBE DMR detection in 1992 of large angular
scale fluctuations attributed to the acausal conditions at the
beginning of the universe was it realized that the elusive,
causally amplified, subhorizon, 
fluctuations on the last scattering surface that seeded the observed
large-scale structure might be within reach once foregrounds could 
properly be accounted for.

A series of experiments  led to multifrequency  measurements that
could eliminate dust and synchrotron galactic foregrounds.
These had plagued much of the earlier work, but  bolometer
sensitivities  improved and microwave receiver noise temperatures were
reduced while 
bandwidths were increased. The backgrounds could be modeled
with sufficiently high sensitivity,
wide spectral frequency and sky coverage.
The definitive measurements came in at the beginning of the new
millennium, 
from  the BOOMERANG
and MAXIMA bolometric balloon experiments and the DASI, VSA and CBI 
mountain-based interferometers. 

The conclusions have revolutionized cosmology.
The hypothesis of structure formation by gravitational instability
of primordial infinitesimal inhomogeneities has been verified.
The flatness of the universe is measured: the first peak occurs at
$\ell =215\pm 10$ and requires $\Omega_0=1.02 \pm 0.06.$ The power
spectrum of the fluctuations is described by a constant index power
law fit by $n=0.96\pm 0.06,$ essentially the scale-invariant
expectation.
The baryon 
abundance is confirmed to be $\Omega_b h^2=0.021 \pm 0.04.$
All of these quantities are cited at 68\% confidence
(e.g. \cite{archeops}, \cite{ruhl}).

\section {Some theory}
Detection of the CMB fluctuations is a triumph of the simplest Big
Bang cosmology: confirmation of a linearly perturbed expanding
universe.  To describe the anisotropies in a Euclidean geometry, one
expands the temperature field on the celestial sphere in  spherical
harmonics
\begin{equation}
{\Delta T\over T}(\hat{n}) = \sum_{\ell=2}^\infty \sum_{m=-\ell}^{\ell}
  a_{\ell m} Y_{\ell m}(\hat{n}) \; .
\end{equation}
By definition the mean value of the $a_{\ell m}$
is zero.

The
correlation function (or $2-$point function) of the temperature field
$C(\theta)$ is   the average of
$\Delta T/T(\hat{n}_1)\Delta T/T(\hat{n}_2)$
across all pairs of points in the sky $(\hat{n}_1,\hat{n}_2)$ separated by an
angle $\theta$ (~$\cos\theta=\hat{n}_1\cdot\hat{n}_2$).
One usually assumes statistical isotropy and  Gaussian
statistics.  One can now write down \cite{WilSil81} 
a  multipole expansion 
\begin{equation}
\label{cleq}
  C(\theta) = {1\over 4\pi} \sum_{\ell} (2\ell+1) C_\ell P_\ell(\cos\theta)
\end{equation}
where $P_\ell(\cos\theta)$ are the Legendre polynomials and the $C_\ell$ are
the multipole moments,
\begin{equation}
  C_\ell = \sum_{m=-\ell}^\ell  a_{\ell m}^{*} a_{\ell m},
\end{equation}
and where 
$\ell \propto \theta^{-1}$.

Each $C_\ell$ comes from averaging over $2\ell+1$ modes, and  the
sample variance error on $C_\ell$ is
\begin{equation}
  {\delta C_\ell\over C_\ell} = \sqrt{2\over 2\ell+1}f_{\rm sky}^{-1/2}
\end{equation}
where a fraction
$f_{\rm sky}$ of the sky is observed \cite{ssw}, \cite{jungman}.
The contribution to the logarithmic power is usually written as 
${\ell} (\ell+1) C_\ell$, this combination being  constant at low ${\ell}$ for
scale-invariant fluctuations.

The radiation power spectrum shows two characteristic angular scales.
A prominent peak occurs at $\ell\approx 220$ or about 30 arc-minutes. This is
the angular scale that corresponds to the horizon at the moment of
last scattering of the radiation. 
The corresponding comoving scale, corresponding to  the wavelength of
a wave that
just spans  this horizon, is approximately 100 Mpc.
The second scale is the damping scale of about 6 arc-minutes,
 where the finite thickness of
the last scattering surface just spans a wave of comoving scale around
10
Mpc and damps out temperature fluctuations on smaller scales \cite{sdamp}.

The principal features  that separate these scales are the Sachs-Wolfe
plateau at low $\ell,$ the series of peaks between $\ell
 =200-2000$, and the damping tail. The peaks are the relic of coupled
 photon-baryon oscillations, in effect sound waves in the primordial
 plasma, at the epoch of last scattering of radiation  and matter.
One finds a series of peaks because the growing mode of fluctuations is
the only surviving mode, and are phase-locked on super-horizon scales
most likely due to their inflationary origin. Waves that just crest at
last scattering produce the first adiabatic  peak in the CMB. Waves of
half the last horizon  scale also crest, as do smaller waves until
damping sets in at about 10 percent of the wavelength of the first
acoustic peak. The peaks in the power spectrum
are associated with  successive maxima and minima of density
waves in the coupled  photon-baryon plasma, alternately in and out of
phase with 
the dark matter potential wells that dominate the gravity and
correspond to compressions and rarefactions on the last scattering
surface that in turn generate hot spots and cold spots in the
radiation on the last scattering surface.
In quadrature, these are represented as a series of peaks in the power
spectrum.

Coherent peaks are generated  because an inflationary,
or more generally, a very early, origin means that only the growing
mode of fluctuations are present, and since no growth occurs on
superhorizon scales, the phases of the waves can only evolve
once growth commences  after entering
the horizon. Hence for example any wave that just spans the horizon
at last scattering has the same phase.
Any wave that spans half the horizon at last scattering has had
exactly the same amount of subhorizon growth, so these too are in
phase.  Now the wave amplitude  is proportional to
$cos(kt/a),$ where $k/a$ is the  wave-number, $a(t)$ is the 
cosmological scale factor,
and the initial conditions  are  such that the amplitude must be  constant
(and independent of wave-number for a scale-invariant spectrum) on
super-horizon scales $2\pi a/k\gg ct.$ This is the Sachs-Wolfe effect.
 The wave cresting at last scattering $t_{LS}$
gives the first peak,  and there is a series of
compression/rarefaction peaks at $kt_{LS}/a=(n+1)\pi, \ \
n=0,1,2....,$
 and Doppler peaks at $kt_{LS}/a=(n+1/2)\pi, \ \
n=0,1,2.....$ 
The compression mode is  found  to be  dominant.

The Doppler effect \cite{sunyaev}
generates smaller peaks that are 90 degrees out
of phase with the compression mode, and the effect of enhancing the
baryon component is to strengthen the compressions relative to the
rarefactions because of the baryonic contribution to the
gravity \cite{hu}. The baryon density controls the 
rarefaction amplitude  relative to the
compression amplitude, so the odd-even alternation of peak strengths
enables $\Omega_bh^2$ to be read off the sky.

The influence on 
the radiation power spectrum of
spatial curvature  with accurate  matter-radiation coupling
was first computed in 1981 \cite{SilWil81}.
In  curved space, a generalised multipole expansion is necessary.
One finds that spatial curvature modifies  the angular distribution
but its effects are  readily disentangled.
In the modern language of acoustic peaks,   the peaks
are found to be shifted by a factor of order $\Omega^{1/2}$ towards smaller
angular scales in hyperbolic spaces, for progressively lower
$\Omega$ \cite {GouSugSas91}, \cite {KamSpeSug94}). 
Confirmation of the flatness of the universe via measurement of the
peak location represents a major triumph of inflationary cosmology.

There are many complications, some welcome, some less so.  
The overall amplitude 
of the peaks is largely controlled by  $\Omega_m$, or more precisely
the ratio of matter to radiation densities and by $\Omega_\Lambda.$
There is a
degeneracy between $\Omega_m$ and $\Omega_\Lambda$ that cannot be
resolved
by the CMB alone \cite{bond}.
Reionization suppresses the
total power, and in particular all of the peak heights, and can be
constrained \cite{SugSilVit93}. In fact,  $\Omega_\Lambda$   results in
lensing of the peaks and  redistributes power to small scales, so only
very
high $\ell$ measurements can help break this degeneracy 
\cite{MetSil97}
In practice, the degeneracy is best broken by correlation with deep
redshift surveys since the effects of $\Omega_\Lambda$  are most
important at $z\simlt 1$ \cite{seljak}.

\section {Implications}

One frontier has recently
 been breached in the CMB. The first detection of
polarization has been reported by the DASI interferometer. The
polarization in an FRW universe 
 is due to the anisotropy of the
Thompson scattering cross-section combined with the intrinsic
quadrupole component of the radiation fluctuations \cite{kaiser}. 
This results in polarisation
that typically peaks at the angular scale of the last scattering
surface.
Detection  provides additional confirmation of
the standard model.

The cosmic microwave background fluctuations provide an important
probe of inflation. This arises in two ways. Firstly fluctuations on
angular scales larger than a degree are acausal at last scattering
and so are generated at GUT scales during inflation or even earlier as
part of Planck era physics.  One is directly viewing the impact of
quantum fluctuations. Secondly, gravity waves  are generically
generated as a tensor mode in inflationary models. These contribute
only
to
the large angular scale anisotropies, since the tensor mode is
incompressible and redshifts away on subhorizon scales once the
universe is matter-dominated \cite{dautcourt}. 
One can constrain the tensor/scalar
ratio  from the ratio of the low $\ell$ amplitude to the peak heights,
combined with the 2DF power spectrum, 
to be less than about 70 percent \cite{percival02}.  

Polarisation of the temperature
fluctuations will eventually provide   a much more constraining probe
of  both the ionization history  and primordial gravity 
waves \cite{polnarev}. In general, the polarisation can be separated into
two modes that can be expressed  as the gradient and the curl of a potential. 
Electron scattering of a radiation field with an intrinsic quadrupole
component generates  the gradient mode of polarisation. Gravitational
lensing couples this gradient to the curl mode. However only 
gravity waves give an intrinsic curl component of polarisation. This
provides a unique signature that can in principle be disentangled from
the
induced curl polarisation, and should be detectable on degree scales or
larger at the 0.5$\mu$K level for typical inflationary models \cite{SelZal}.

Reionisation further complicates matters, since late electron scattering
also polarises the CMB. For typical models, 
the detailed $C_\ell$
spectrum of scalar-induced polarisation should be  distinguishable at a
 level of 5--10$\mu$K on small angular scales and around 1$\mu$K on large scales
$(\ell \simlt 200)$  \cite{seljakpol}. Late reionization 
produces  the gradient mode of polarisation and has a
characterisic low $\ell$ bump corresponding to the horizon size 
at reionization \cite{hw}, \cite{kap}.
In fact the gradient terms contribute to polarisation 
at a level of about $0.1\delta T/T,$ and this has  been detected
by the DASI interferometer on 5-30 arc-minute scales \cite{kovac}. 
The MAP satellite
should be capable of seeing the characteristic peak associated  with
reionization, if this occurred at $z\simgt 10.$  

The tensor
contribution to the curl component of polarisation is only of order
$0.01\delta T/T,$ and detection will require dedicated experiments that
must await the post-Planck era, 2008 or later.
Detection of the gravity wave mode would
provide another  powerful confirmation of inflation, which generically
predicts a mixture of scalar and tensor relic  fluctuation modes
with approximately scale-invariant spectra.

In contrast, pre-Big Bang and ekpyrotic models usually predict blue spectra of
primordial gravity waves that would not leave any detectable CMB relic
 polarisation. Intrinsic non-gaussianity is an aspect of generic very
 early universe models  that could also help constrain
 inflation. There are hundreds  of variants of inflationary cosmology,
 and it is difficult to come up with any definitive tests,
other than perhaps the prediction of flatness of space. Multiple
 scalar fields that drive several phases of inflation, or interact
 during the primary phase, will lead to  non-gaussian
 fluctuations. Topological defects such as cosmic strings or textures
 that are generated  during post-inflationary reheating are a promising
 source of non-gaussian fluctuations on small angular scales,
although these must be subdominant with regard to structure formation.

At high $\ell,$ the observed damping provides an independent probe  of
$\Omega_b,$  and confirmation of the physics of last scattering.
Beyond $\ell\sim 2000,$ non-linear effects play an important role.
The most prominent of these is the Sunyaev-Zeldovich (SZ) effect, which is
the cumulative contribution of hot gas in distant 
unresolved clusters along the line of
sight,
 and is uniquely
distinguishable by its spectral characteristics as an apparent
deficit below 218 GHz in the rest frame 
and excess at higher frequency, relative to the 
CMB. There is a reported 2$\sigma$ detection at $\ell\sim 2500$
by the CBI experiment which however only covers a narrow frequency range near 30
Ghz \cite{mason}.

The Ostriker-Vishniac \cite{ov}
and Rees-Sciama \cite{rs} effects are also of interest at
$\ell\simgt 3000,$   due respectively to correlations of  linear 
compressions and rarefactions with 
Doppler contributions along the line of sight  and to traversal by CMB
photons across  the time-dependent potential wells associated with
forming galaxy clusters, respectively.
 These anisotropies  are expected
to amount to only a percent or so of the SZ effect, as also is the
kinematic SZ effect. However the challenge of detecting secondary
fluctuations is important, since they probe the epoch of reionisation.
Another aspect that is characteristic of secondary fluctuations is
that they are expected to be intrinsically non-Gaussian, as well as
being anti-correlated with the primary fluctuations.
In general, reionization damping of primary fluctuations generates
secondary fluctuations and enhances the polarisation:
a recent discussion is given in  \cite{Hu00}.

The acausality of large angular scale fluctuations $(\simgt 1^\circ)$
means that they were generated either as initial conditions or by an
early de Sitter phase of inflation. The latter occurred in the first
$10^{-35}$ second, and amplified quantum fluctuations onto macroscopic
scales. An example of a novel approach to initial conditions comes
from the ekpyrotic cosmology. This model
 dispenses with inflation and sets
up the fluctuations in a pre-Big Bang phase that can be identified
with the past history of two colliding branes. The
brane  intersection
signals the initial moment of the Big Bang at the Planck time.
This is also a singularity in space-time, and the calculations of the
emergent density fluctuations are controversial.

The synthesis of baryons occurs at electroweak (100 GeV) or GUT
($10^{16}$ GeV) scales, and is responsible for the specific entropy of
the universe, that is, the number of photons per baryon observed
in the CMB.  
It is tied in with lepton synthesis and generation of lepton number
$L$ , since $B-L$ is generally
conserved in subsequent phase transitions that may dilute any
primordial baryon number $B$.
The blackbody spectrum itself was generated in the first
hours of the Big Bang. Double Compton scattering and bremsstrahlung 
are the key processes responsible for thermalisation. An
essentially  perfect
blackbody is generated. No trace of thermalization processes is left,
other than by possible late energy injection after the thermalization epoch.

The matter and radiation remain coupled
until a redshift of about 1000. Hence the temperature fluctuations
trace the degree of matter inhomogeneity over the first 300,000
years of the universe. One mostly measures the imprint of density fluctuations
on the last scattering surface. These dominate the acoustic
peaks. There is also a contribution at lower $\ell$ from the
integrated Sachs-Wolfe effect, both before and after last scattering,
due to the decay of the ratio of the radiation to the matter densities.

The temperature fluctuation mapping has led to an era of precision
cosmology. One measures in addition to $\Omega_0$ and $n$,  $\Omega_b$ 
as well as a combination of $\Omega_\Lambda$ and $H_0.$ With
independent determinations of $\Omega_m$  from large-scale structure,
the standard model is well defined and well constrained.

\section {Towards the Future}

There are some 10 parameters that describe the CMB fluctuations, in
addition to the 6 or so that characterise the standard moodel of
cosmology.
What is remarkable is that precise measurements of the CMB
fluctuations
can potentially measure almost all of these parameters with unrivalled
precision.
There are degeneracies, most notably between $\Omega_m$ and 
$\Omega_\Lambda,$ but these can largely be resolved by
cross-correlating with other data  such as wide area weak lensing surveys.
These are exciting times for the cosmic microwave background,
especially 
in combination with other data sets.
The precision of the measurements is vastly superior to that of a
decade ago. The overall picture is one of consistency with what is
rapidly emerging as the standard model of cosmology.
 
There are less conventional issues that the CMB may eventually tackle.
Gravity could change on very large scales, and the CMB uniquely probes
the largest scales via the integrated Sachs-Wolfe effect, thereby
providing a potentially unique probe of exotic gravity models \cite{bin}.
The  topology of the universe is not
predicted by general relativity, or for that matter
by quantum gravity.
The CMB may provide the only means of ever detecting a topological
signature of space-time. A flat, compact topology
can undergo inflation from a quantum origin \cite{starob}.
 If the topological scale is on the order of
the horizon scale, a detectable imprint in the form of a non-Gaussian
pattern is generated on the CMB sky \cite{SteScoSil93}.
The existing COBE data eliminate all toroidal topologies if the
topological scale is of order the present  horizon scale or less
\cite{angel}. 
 The universe is locally isotropic
but globally anisotropic in many of the topological representations,
which,  in the case of a flat universe,
 only amount to a handful of more general  possibilities. 
It will be interesting to eventually 
 explore some of the resulting patterns with more sensitivity
and better resolution
\cite{stark}.

One can hope to probe the beginning of the universe;
indeed, one already does this with large angular scale fluctuations.
There is considerable support for inflation from the observed flatness
of the universe, as well as 
from the approximate   scale invariance of the power spectrum
and the coherence of the acoustic peaks, although one hesitates to
cite these latter two  predictions as  
 robust inferences from  a theory which is 
neither unique nor able to account for
the amplitudes of the peaks. The acausality at the epoch of last
scattering on large angular scales could equally be created  in
the quantum chaos of the Planck epoch or even earlier, via pre-Big Bang
physics,
all that is required being  phase coherence on superhorizon scales of
matter and radiation fluctuations. The polarization signal of 
a relic gravity
wave background presents the ultimate challenge, and the ultimate probe.

Most of this is for the future. At present, one has some noteworthy
highlights. Gravitational instability theory
of structure formation has been verified.
The universe has been
found to be spatially flat. 
The CMB  radiation power spectrum has 
provoked the death of topological defects as a
significant  source of seeds for structure formation. 
A purely baryonic universe can be discarded. 
This is remarkable progress in modern cosmology.


\begin{thebibliography}{99}

\bibitem{SacWol67} 
``Perturbations of a Cosmological Model and Angular Variations of the
Microwave Background,''
R.K.  Sachs \& A.M. Wolfe, ApJ {\bf 147}, 73-90 (1967)


\bibitem{s67}
 ``Fluctuations in the Primordial Fireball,''
J. Silk, Nature {\bf 215}, 1155-1156 (1967)





\bibitem{cobe}
 ``Structure in the COBE differential microwave radiometer
 first-year maps,'' G.F.Smoot, {\it et al.}, ApJ {\bf 396}, L1-5 (1992)  





\bibitem{miller}
``A Measurement of the Angular Power Spectrum of the CMB from l = 100
to 400,
'' A. Miller {\it et al.}, ApJ {\bf 524}, L1-L4 (1999)



\bibitem{debernadis}
``A flat Universe from high-resolution maps of the cosmic microwave
background radiation,''
P. de Bernardis  {\it et al.}, Nature, {\bf 404}, 955-959 (2000)

\bibitem{hanany}
``MAXIMA-1: A Measurement of the Cosmic Microwave Background
Anisotropy on Angular Scales of 10$'$ to 5$^\circ$,''
S. Hanany {\it et al.}, ApJ {\bf 545}, L5-L9 (2000)






\bibitem{padin}
``First Intrinsic Anisotropy Observations with the Cosmic Background
Imager,''
S. Padin  {\it et al.}, ApJ {\bf 549}, L1-L5 (2001).

\bibitem{halverson}
``Degree Angular Scale Interferometer First Results: A Measurement of
the Cosmic Microwave Background Angular Power Spectrum,'' N. Halverson 
{\it et al.}, ApJ {\bf 568}, 38-45 (2002)

\bibitem{taylor}
``First results from the Very Small Array -- II. Observations of the CMB,''
A. Taylor {\it et al.}, MNRAS, in press (2002), astro-ph/0205381
\bibitem{archeops}
``The Cosmic Microwave Background Anisotropy Power Spectrum measured
by Archeops,''
A. Benoit {\it et al.}, A\&A, in press (2002), astro-ph/0210305

\bibitem{lifshitz}
``On the Gravitational Instability of the Expanding Universe,''
E. Lifshitz, 
Journal of Physics USSR, {\bf 10},  116-122 (1946)

\bibitem{wss}
``From Microwave Anisotropies to Cosmology,'' M. White, J. Silk
and D. Scott,  Science,  {\bf 268}, 829- 835(1995)


\bibitem{gawiser}
``Extracting Primordial Density Fluctuations,'' E. Gawiser and
J. Silk, Science, {\bf 280}, 1405-1411 (1998)




\bibitem{tegmark}
``Separating the Early Universe from the Late Universe: cosmological
parameter estimation beyond the black box,'' M. Tegmark and
M.Zaldiarraga,
PRD, in press (2002).

 
\bibitem{peacock}
``A measurement of the cosmological mass density from clustering in the
2dF Galaxy Redshift Survey,''
J. Peacock et al., Nature,  {\bf 410}, 169-173 (2001)

\bibitem{lahav}
``The 2dF Galaxy Redshift Survey: the amplitudes of fluctuations in
the 2dFGRS and the CMB, and implications for galaxy biasing,''
O. Lahav {\it et al.}, MNRAS,  {\bf 333}, 961-968 (2002)


\bibitem{brown}
``The shear power spectrum from the COMBO-17 survey,'' M. Brown {\it et al.},
 MNRAS, submitted, astro-ph/0210213 (2002)

\bibitem{hoekstra}
``Weak lensing study of galaxy biasing,''
H.Hoekstra, L. van Waerbeke, M. Gladders, Y. Mellier and H. Yee,
ApJ, in press (2002),  astro-ph/0206103

\bibitem{schuecker}
``The REFLEX Galaxy Cluster Survey VII: $\Omega_m$ and $\sigma_8$ from cluster
abundance and large-scale clustering,''
P. Schuecker, H. Boehringer, C. Collins and L. Guzzo, A\&A, in press
(2002),
astro-ph/ 0208251

\bibitem{gamow}
``Expanding Universe and the Origin of the Elements,''
G. Gamow, Phys. Rev.  {\bf 70}, 572-573 (1946)
\bibitem{alpher}
``Evolution of the Universe,''
R. Alpher and R. Herman, Nature  {\bf 162}, 774-775 (1948)

\bibitem{zeld}
Y. Zeldovich, JETP, {\bf 43}, 1561 (1962)

\bibitem{dorosh} 
``Mean Radiation Density in the Metagalaxy and Some Problems of
Relativistic Cosmology,'' A. G. Doroshkevich and I. D. Novikov,
Doklady Akadamie Nauk. USSR,  {\bf 154}, 809--811 (1964)


\bibitem{mather} 
``Calibrator Design for the COBE  Far-Infrared Absolute
  Spectrophotometer (FIRAS),''
J.C. Mather, D.J. Fixsen, R.A. Shafer, Mosier and Wilkinson, D.,
 ApJ {\bf 512}, 
511-520,  (1998) 

\bibitem{harrison}
``Fluctuations at the Threshold of Classical Cosmology,''
E.Harrison, PRD  {\bf 1}, 2726-2730 (1970)




\bibitem{PY}
``Primeval Adiabatic Perturbation in an Expanding Universe,''
 P.J.E. Peebles \& J.T. Yu,
ApJ {\bf 162}, 815-836  (1970) 

\bibitem{zeldovich}
``A Hypothesis Unifying the Structure and Entropy of the Universe,''
Y. Zeldovich, MNRAS {\bf 160}, 1P-3P (1972)


\bibitem{guth}
``Inflationary Universe: a Possible Solution to the Horizon
 and Flatness Problems,'' A. Guth,  PRD  {\bf 23}, 347-356 (1981) 

\bibitem{linde}
``A New Inflationary Universe Scenario:  a Possible Solution of the Horizon,
  Flatness, Homogeneity,  Isotropy and Primordial Monopole Problems,''
 A Linde,  Phys. Lett., {\bf 108B}, 389-393 (1982)

\bibitem{albrecht}
``Cosmology for grand unified theories 
with radiatively induced symmetry breaking,''
A. Albrecht \& P.  Steinhardt, PRL  {\bf 48}, 1220-1223 (1982)
\bibitem{pee68}
``Recombination of the primeval plasma'', P.J.E. Peebles,
ApJ, 153, 1-12 (1968)  



\bibitem{BonEfs84}
``Cosmic background radiation anisotropies in universes 
dominated by nonbaryonic dark matter,''
J.R. Bond \& G. Efstathiou, ApJ {\bf 285}, L45-48 (1984) 

\bibitem{VitSil84}
``Fine scale anisotropy of the cosmic microwave background in a universe
dominated by cold dark matter'',
N. Vittorio, J. Silk, ApJ, 285, L39-43 (1984) 


\bibitem{DSZ}
``Fluctuations of the microwave background radiation in the 
adiabatic and entropic theories of galaxy formation,''
A.G. Doroshkevich, I.B. Zel'dovich, R.A. Sunyaev,
Soviet Astronomy {\bf 22}, 523-528 (1978) 


\bibitem{SilWil80}
 ``Residual Fluctuations in the Matter and Radiation Distribution after
the Decoupling Epoch,'' J. Silk and M. L. Wilson,
Physica Scripta, {\bf  21},
708-713 (1980).




\bibitem{sunyaev}
``Small-Scale Fluctuations of Relic Radiation,''
R. Sunyaev and Ya. B. Zeldovich, 
Ap\&SS,   {\bf 7}, 3-19 (1970)

\bibitem{sak}
A. Sakharov, JETP {\bf 49}, 345 (1965)

\bibitem{percival}
``The 2dF Galaxy Redshift Survey: 
The power spectrum and the matter content of the universe,''
W. Percival {\it et al.}, MNRAS, {\bf 327}, 1297-1306 (2001)
\bibitem{ruhl}
``Improved Measurement of the 
Angular Power Spectrum of Temperature Anisotropy in the
CMB from Two New Analyses of BOOMERANG Observations,'' J. Ruhl
{\it et al.}, astro-ph/0212229 (2002)

\bibitem{WilSil81}
``On the anisotropy of the cosmological background matter and radiation 
distribution. I - The radiation anisotropy in a spatially flat universe,''
M.L. Wilson \& J. Silk, ApJ {\bf 243}, 14-25 (1981) 

\bibitem{ssw}
``Sample variance
 in small-scale cosmic microwave background anisotropy experiments,''
 D. Scott, M. Srednicki and M. White, ApJ {\bf 421}, L5-L7 (1994)

\bibitem{jungman}
``Cosmological-parameter determination with microwave background maps,''
G. Jungman,  M. Kamionkowski, A. Kosowsky and D. Spergel,
Phys. Rev. D  {\bf 54}, 1332-1344 (1996)

\bibitem{sdamp}
``Cosmic Black-Body Radiation and Galaxy Formation,'' J. Silk, 
ApJ {\bf 151}, 459-472 (1968)   


\bibitem{hu} 
``Toward understanding CMB anisotropies and their implications,''
W. Hu \& N. Sugiyama,  PRD {\bf 51}, 2599-2630, (1995) 

\bibitem{SilWil81}
``Large-scale anisotropy of the cosmic microwave background radiation,''
J. Silk \& M.L. Wilson,  ApJ {\bf 244}, L37-42 (1981) 
 
\bibitem{GouSugSas91}
``Large Angle Anisotropy of the Cosmic Microwave Background in an Open 
Universe,''
N. Gouda, N. Sugiyama, M. Sasaki, Prog Theor Phys {\bf 85}, 1023-1039 (1991) 


\bibitem{KamSpeSug94}``Small-Scale Cosmic Microwave Background Anisotropies 
as a Probe of the Geometry of the Universe,''
M. Kamionkowski, D. N. Spergel, N. Sugiyama,
ApJ {\bf 426}, L57-60   (1994) 




\bibitem{bond}
``Measuring cosmological parameters with cosmic 
microwave background experiments,''
J. Bond  {\it et al.}, PRL, {\bf 72}, 13-16 (1994)

\bibitem{SugSilVit93}
``Reionization and Cosmic Microwave Anisotropies,''
N. Sugiyama, J. Silk, N. Vittorio, ApJ {\bf 419}, L1-4 (1993) 




\bibitem{MetSil97}
``Gravitational Magnification of the Cosmic Microwave Background,''
R.B. Metcalf \& J. Silk,
ApJ {\bf 489}, 1-6, (1997)  



\bibitem{seljak}
``Measuring Dark Matter Power Spectrum from Cosmic Microwave Background,''
U. Seljak \& M. Zaldiarraga, PRL, {\bf 82}, 2636-2639 (1999)


 \bibitem{kaiser}
``Small-angle anisotropy of the microwave background radiation in the
adiabatic theory,'' N. Kaiser, 
MNRAS, {\bf 202}, 1169-1180 (1983)


\bibitem{dautcourt}
``Small-scale variations in the cosmic microwave
  background,'' G.  Dautcourt, MNRAS, {\bf 144}, 255-278  (1969)

\bibitem{polnarev}
``Polarization and Anisotropy Induced in the Microwave Background by
Cosmological Gravitational Waves'',
A.G. Polnarev, Soviet Astronomy, {\bf 29}, 607-613 (1985)  

\bibitem{percival02}
``Parameter constraints for flat cosmologies from CMB and 2dFGRS power
 spectra,''
W. Percival {\it et al.}, MNRAS, in press, 
astro-ph/0206256
\bibitem{SelZal}
``Signature of Gravity Waves in Polarization of the Microwave Background,''
U. Seljak \& M. Zaldarriaga, Phys.Rev.Lett. {\bf 78}, 2054-2057,
astro-ph/9609169 (1997)  


\bibitem{seljakpol}
``Measuring Polarization in the Cosmic Microwave Background,''
U. Seljak, ApJ {\bf 482}, 6-16 (1997)

\bibitem{hw}
``The Damping Tail of Cosmic Microwave Background Anisotropies,''
 W. Hu and M. White,  ApJ {\bf 479}, 568-579 (1997)
\bibitem{kap}
 ``Probing the Reionization History of the Universe using the Cosmic Microwave Background Polarization,''
 M. Kaplinghat, M. Chu, Z. Haiman, G. Holder, L. Knox, C. Skordis,
ApJ, in press (2002)  astro-ph/0207591

\bibitem{kovac}
 ``Detection of Polarization in the Cosmic Microwave Background using
 DASI,''
 J. Kovac, E. M. Leitch, C. Pryke, J. E. Carlstrom, N. W. Halverson 
and  W. L. Holzapfel, in press (2002)  astro-ph//0209478






\bibitem{mason}
 ``The Anisotropy of the Microwave Background to l = 3500: Deep Field Observations with the Cosmic
Background Imager,'' B. Mason {\it et al.},  ApJ, in press (2002)
  astro-ph//0205384


 
\bibitem{ov}
``Generation of microwave background fluctuations from 
nonlinear perturbations at the era of galaxy formation,'' 
J. Ostriker \& E. Vishniac,  ApJ {\bf 306}, L51-54 (1986)  

\bibitem{rs}
``Large-scale density inhomogeneities in the universe''
M. J. Rees \& D.W. Sciama, Nature {\bf 217}, 511-516 (1968) 




\bibitem{Hu00}
``Reionization revisited: secondary CMB anisotropies and polarization,''
W. Hu, ApJ {\bf 529}, 12-25, astro-ph/9907103  (2000) 


\bibitem{bin}
``Probing Large Distance Higher-Dimensional Gravity with Cosmic Microwave
 Background Measurements,'' P. Binetruy and J. Silk,
PRL {\bf 87},
 031102(1)-031102(4) (2001)



\bibitem{starob}
``Quantum Creation of a Universe with Nontrivial Topology,''
Ya. Zeldovich \& A.  Starobinskii, 
Soviet Astr. Lett., {\bf 10}, 135-137 (1984)


\bibitem{SteScoSil93}
``Microwave background anisotropy in a toroidal universe,''
D. Stevens, D. Scott, J. Silk, PRL {\bf 71} 20-23, (1993)  



\bibitem{angel}
``Can the Lack of Symmetry in the COBE DMR Maps Constrain the Topology
of the Universe?,''
A. Oliveira-Costa, G.Smoot and A. Starobinsky, ApJ,
{\bf 468}, 457-461 (1996) 

\bibitem{stark}
``Circles in the Sky: Finding Topology with the 
Microwave Background Radiation,''
N. Cornish, D. Spergel, G. Starkman,
Class.Quant.Grav. {\bf 15},  2657-2670 (1998)  

\end{thebibliography}
\end{document}